\documentclass[11pt]{article}
\usepackage{amsfonts,amssymb,latexsym}
\title{\bf String with noncommutative world-sheet\\ and stringy instantons}
\author{I.V.Gorbunov\footnote{}\ \
and A.A.Sharapov\footnote{}\\
{\it Department of Physics, Tomsk State University,}\\
{\it Lenin Ave. 36, Tomsk, 634050 Russia}}
\date{\bf hep-th/0201033}
\begin{document}
\maketitle
\renewcommand{\thefootnote}{\fnsymbol{footnote}}%
\footnotetext[1]{{\it e-mail:} ivan@phys.tsu.ru}%
\footnotetext[2]{{\it e-mail:} sharapov@phys.tsu.ru}%
\renewcommand{\thefootnote}{\arabic{footnote}}%
\setcounter{footnote}{0}%
\begin{abstract}
The model of the bosonic string with the noncommutative
world-sheet geometry is proposed in the framework of Fedosov's
deformation quantization. The re\--inter\-pre\-ta\-tion of the
model in terms of bosonic string coupled to infinite multiplet of
background fields is given and the link to the $W$-symmetry is
discussed. For the case of d=4 Euclidean target space, the
stringy counterparts of the Yang-Mills instantons are constructed
in both commutative and noncommutative regimes.\\[1mm]
\end{abstract}
\section{Introduction}\hspace*{\parindent}
The passage from the paradigm of point particles to strings inevitably
introduces a new fundamental scale related to the string size and
leads to a new type of ``fuzziness'' in addition to the
quantum-mechanical uncertainty. The quantum effects are
proportional to the Planck constant $\hbar$, while the stringy
effects are governed by the reciprocal of string tension $\alpha'$
and both of these parameters are responsible for a certain
deformation of the classical phase-space and/or target-space
geometry underlying the string theory. Although there is a strong
evidence that varying  these parameters one may interpolate
between five known string models and even 11-dimensional
supergravity, a constructive definition of the full theory in the
``intermediate region", called M-theory, remains a mystery. As to
the $\hbar$-deformation, we have the powerful machinery of modern
quantization methods allowing to handle efficiently the problem in
many cases. (It is the requirement of quantum consistency that
appears to be a fundamental vehicle for identifying the five
string theories as appropriate points on $(\hbar,\alpha')$-scan
of M-theory.) Meanwhile, the mathematical structure behind the
$\alpha'$-deformation is less well understood.

Nowadays most of the hope for obtaining a microscopic description
of M-theory centers around the M(atrix) theory
\cite{BFSS97,Taylor00}. Perhaps, the simplest way to see how
the Matrix models are related  to string theory is to start with
the ``Poisson formulation" of the bosonic string~\cite{IKKT,CDS97},
\begin{equation}
S=\int\limits_M\omega\left(\{X^A,X^B\}^2-\alpha^2\right)\,.
\label{i1}
\end{equation}
Here $X^A$ are the target-space coordinates and
$\{\,\cdot,\,\cdot\}$ stands for the Poisson bracket associated
to the area two-form $\omega$ on the string world-sheet $M$. At
the level of classical equations of motion the model (\ref{i1}) is
equivalent to the Nambu-Goto string with
$\alpha^\prime=\alpha/2\pi$. The Matrix theory approach is based
on exploiting a similarity between algebraic properties of the
Poisson bracket and the matrix commutator. Considering the
coordinates $X^A$ to be Hermitian  $N\times N$-matrices and making
a formal substitutions
\begin{equation}
\{\,\cdot\,,\,\cdot\}\Rightarrow
i[\,\cdot\,,\,\cdot\,]\,,\qquad \int\limits_M\omega\Rightarrow{\rm
Tr}\,, \label{i2}
\end{equation}
in the action (\ref{i1}) one comes to what is known as IKKT
matrix model\footnote{More precisely, we consider only bosonic sector
of this model.}\ \cite{IKKT}:
\begin{equation}
S_{\rm IKKT}=-{\rm Tr}\,\left( [X^A,X^B]^2+\alpha^2\right)\,. \label{i3}
\end{equation}
Up to the constant term this action function can be viewed as
the result of dimensional reduction to a point of ${\rm U}(N)$ Yang-Mills
theory. It is assumed that taking a certain large-$N$ limit of
this model one can approximate the bosonic string action (\ref{i1}) provided
the eigenvalue distributions of the operators $X^A$ are
smooth enough. For two-torus and sphere such a limit can be traced
explicitly, see e.g.~\cite{Taylor00}.

It is very fruitful in physics to treat the operators as not to be
given \textit{a priori} but resulting from the quantization of
phase-space observables. Assuming the  IKKT model to be somehow
related to the string theory it is more than natural to consider
a quantization of the Poisson bracket entering to the action
(\ref{i1}). As we will show bellow this can be actually done
within the framework of Fedosov's deformation quantization
\cite{Fedosov,Fedosovbook}. What we get as a result may be
regarded as the string with noncommutative world-sheet. From a
general viewpoint this approach, as such, is very similar to that
of noncommutative field theory, if not identical in two respects.
First, the noncommutativity of geometry underlying our model is
\textit{dynamical}, i.e. governed by the unconstrained field
$\omega$, in contrast to a constant Poisson bracket used  in the
known models of noncommutative field theory. The second point,
closely related to the first one, is that consistent definition
of the star-product requires introducing a \textit{symplectic
connection} respecting form~$\omega$. Unlike the metric
connection,  the symplectic one is not completely determined in
terms of the structure it preserves, i.e. via the form $\omega$,
and thereby it should be considered as an additional dynamical
field of the theory. The important observation behind our
construction is that both prerequisites of the Fedosov
quantization -- symplectic structure and connection -- are
already contained in the string theory in a form of the Polyakov
metric on the string world-sheet. Namely, we identify $\omega$
with the Riemannian volume-form, which is obviously preserved by
the metric connection.

In this respect, the Deformation
quantization allows us to link the Matrix and Noncommutative field
theories in the quite natural way. The details of the construction
are presented in the Section~2. There we also show that the model
admits a straightforward re-interpretation as the conventional
bosonic string coupled to a special background of external
fields. This background is recognized as the one compatible with
$W$-type symmetry generated by internal star-automorphisms of the
theory. In so doing, the deformation parameter is naturally
identified with the fundamental stringy scale $\alpha'$.

Returning to the question about the nature of
$\alpha'$-deformation we may say that, in certain respects, it can
be put in the same mathematical framework as $\hbar$-deformation,
i.e. in the framework of Deformation quantization theory.

In Section 3 we address the problem of constructing exact
solutions to our model. On the whole, the situation here closely
resembles the present state of the art of finding explicit
solutions to the YM equations: for arbitrary dimension and
signature the problem looks hopeless, but in special case of
four-dimensional Euclidean space the series of instanton
solutions can explicitly be constructed. Using the similarity
between the forms of IKKT and YM action functionals we derive the
stringy counterpart for the self-duality condition and then
resolve it in a purely algebraic manner.  In the commutative
limit these solutions pass to the holomorphic curves in
$\mathbb{C}^{\,2}\sim\mathbb{R}^{\,4}$ which seem to exhaust all globally
minimal two-surfaces in $\mathbb{R}^{\,4}$. It should be noted,
however, that contrary to the case of YM instantons these are not
localized on the world-sheet in both commutative and noncommutative
regimes.
The reason is that we are confined to solutions with analytical
dependence on the deformation parameter. The existence of
nonanalytical solutions remains an open question and we conclude
with some remarks upon a possible line of attack on the problem.
\section{Deformation of bosonic string}\hspace*{\parindent}
Consider the world-sheet of bosonic string as a Riemannian manifold $M$ with
a metric~$g_{ij}$, $i,j=1,2$. Being a two-dimensional manifold, $M$
also carries a symplectic structure given by the canonical volume
form $\omega=\sqrt{g}\,du\wedge dv$, $g=|{\rm det}( g_{ij})|$. In
fact, $M$ is K\"ahlerian manifold and both structures can be combined
to one degenerate Hermitian form
\begin{equation}
\Lambda^{ij}=g^{ij}+\sqrt{-1}\omega^{ij}\,,
\label{1}
\end{equation}
where $g^{ij}$ and $\omega^{ij}$ are the inverse Riemannian and
symplectic metrics, respectively. The right and left null-vectors
of~$\Lambda$ can be visualized as those spanned light-cone at
each point of~$M$.

In the conventional string theory the coordinates of target space
$X^A(\sigma,\tau)$ are considered to be bosonic fields
forming a commutative algebra with respect to the ordinary pointwise
multiplication of functions on $M$. The noncommutative generalization
of the theory is obtained by deforming the algebra of classical
fields. More precisely, we introduce a new associative operation, called
star-product, being an one-parametric deformation of the pointwise one.
This $\ast$-product is made of Hermitian structure~(\ref{1}) and its first
several terms are given by
\begin{equation}
\begin{array}{c}\displaystyle\!\!\!\!
X^A\ast X^B=X^A\cdot X^B+\frac{\nu}{2}\Lambda^{ij}\nabla_iX^A
\nabla_jX^B+\frac{\nu^2}{8}\Lambda^{i_1j_1}\Lambda^{i_2j_2}\nabla_{i_1}
\nabla_{i_2}X^A\nabla_{j_1}\nabla_{j_2}X^B\left(1-\frac{\nu}{4}R\right)+
\\ \displaystyle+\frac{\nu^3}{48}\Lambda^{i_1j_1}
\Lambda^{i_2j_2}\Lambda^{i_3j_3}\nabla_{i_1}\nabla_{i_2}\nabla_{i_3}X^A
\nabla_{j_1}\nabla_{j_2}\nabla_{j_3}X^B+O(\nu^4)\,,
\end{array}
\label{2}
\end{equation}
where $\nu$ is a deformation parameter, $\nabla_i$ and $R$ are
the covariant derivative and the scalar curvature associated to the
metric $g_{ij}$. The higher terms $O(\nu^4)$ are restored
step-by-step from an associativity condition for the
$\ast$-product. The explicit recurrent formulas can be found in
\cite{Bord,Dolg,Kar}.

Generally speaking, there are many other star-products associated
to the same metric.  For example, in Eq.~(\ref{1}) tensors
$\omega$ and $g$ can be combined with arbitrary complex
coefficients. The above $\ast$-product corresponds to a nonlinear
generalization of the Wick symbol associated with the usual
creation-annihilation operators. The use of the Wick star-product
is seemed to be natural in the context of string theory since it
implies, for example, the commutativity of left (right) modes of
the closed string. More precisely, the product of functions
$X^A(x^\pm)$ depending on one of curvilinear light-cone
coordinates~$x^\pm(u,v)$,
\begin{equation}
\Lambda^{ij}\partial_jx^+=\partial_jx^-\Lambda^{ji}=0\,,
\label{light-cone}
\end{equation}
does not deformed since $x^\pm\ast x^\pm=x^\pm\cdot x^\pm$ for the Wick
$\ast$-product (\ref{2}).

To define the deformed string action we need one more ingredient
of the deformation quantization -- the trace functional on
$\ast$-algebra. It is defined as any linear functional vanishing
on the star-commutator of functions
\begin{equation}
\begin{array}{ll}\displaystyle
[X^A,X^B]&\displaystyle\equiv X^A\ast X^B-X^B\ast X^A=\\&
\displaystyle=\nu\{X^A,X^B\}+\frac{\nu^2}{2}
\omega^{i_1j_1}g^{i_2j_2}\nabla_{i_1}\nabla_{i_2}X^A\nabla_{j_1}\nabla_{j_2}
X^B\left(1-\frac{\nu}{4}R\right)+\\& \displaystyle
+\frac{\nu^3}{48}
\left(\Lambda^{i_1j_1}\Lambda^{i_2j_2}\Lambda^{i_3j_3}-
\Lambda^{j_1i_1}\Lambda^{j_2i_2}\Lambda^{j_3i_3}\right)
\nabla_{i_1}\nabla_{i_2}\nabla_{i_3}X^A
\nabla_{j_1}\nabla_{j_2}\nabla_{j_3}X^B+O(\nu^4)\,.
\end{array}
\label{3}
\end{equation}
The trace functional is known to exist for any symplectic manifold
and can be represented in the form
\begin{equation}
{\rm Tr}\,f=\int\limits_M{\rm d}\mu\,f\,,\qquad\qquad
{\rm Tr}\,[f,g]=0\,,
\label{4}
\end{equation}
where the trace-measure ${\rm d}\mu$ is uniquely determined up to
an overall constant. As is shown in Ref.~\cite{Fedosovbook}, the
measure has a form of  series in $\nu$ with coefficients depending
on the curvature tensor and its derivatives. In our case this
series starts with
\begin{equation}
{\rm d}\mu=\omega\left(1-\frac{\nu}{4}R+\frac{\nu^2}{24}\Box R+O(\nu^3)
\right)\,,
\label{5}
\end{equation}
$\Box$ being the two-dimensional Laplace operator.

The action functional for the deformed string can now be written as
\begin{equation}\label{af}
\begin{array}{l}
\displaystyle
\qquad\qquad S_{def}=\frac{1}{\nu^2}\int {\rm
d}\mu\left([X^A,X^B]\ast[X_A,X_B]-\nu^2\alpha^2\right)= \\[7mm]
\displaystyle\qquad
=\int\omega\left(\{X^A,X^B\}^2-\alpha^2+\frac{\nu}{4}R\{X^A,X^B\}^2
+\nu\{X_A,\{X^B,X^B\}\}\Box X_B\right.+\\ \displaystyle\qquad
+\nu^2\left[\frac{1}{12}\Lambda^{i_1j_1}\Lambda^{i_2j_2}\Lambda^{i_3j_3}
\nabla_{i_1}\nabla_{i_2}\nabla_{i_3}X^A\nabla_{j_1}\nabla_{j_2}\nabla_{j_3}
X^B\{X_A,X_B\}+\right.\\ \displaystyle\!\!\!\!\!
+\frac{1}{4}(\omega^{i_1j_1}g^{i_2j_2}
\nabla_{i_1}\nabla_{i_2}X^A\nabla_{j_1}\nabla_{j_2}X^B)^2-
\frac{R}{8}\Lambda^{i_1j_1}\Lambda^{i_2j_2}\nabla_{i_1}
\nabla_{i_2}X^A\nabla_{j_1}\nabla_{j_2}X^B\{X_A,X_B\}+\\[3mm]
\displaystyle\qquad
+\frac{1}{8}\Lambda^{i_1j_1}\Lambda^{i_2j_2}\nabla_{i_1}\nabla_{i_2}\{X^A,X^B\}
\nabla_{j_1}\nabla_{j_2}\{X_A,X_B\}-\\[3mm] \displaystyle\qquad-
\frac{1}{4}\Lambda^{i_1j_1}\Lambda^{i_2j_2}\nabla_{i_1}\nabla_{i_2}X^A
\nabla_{j_1}\nabla_{j_2}X^B\Box\{X_A,X_B\}+\\[3mm]
\displaystyle\qquad\left.\left.
-\frac{R}{8}\Box\{X^A,X^B\}\{X_A,X_B\}+
\frac{1}{24}\Box R(\{X^A,X^B\}^2-\alpha^2)
\right]\right)+O(\nu^3)\,.
\end{array}
\end{equation}
Besides the global Poincar\'e invariance, the action is invariant, by
construction, under world-sheet diffeomorphisms
\begin{equation}\label{dif}
\delta X^A= \xi^i\partial_iX^A\,,\qquad \delta g_{ij}=
\xi^k\partial_k g_{ij}+g_{kj}\partial_i\xi^k+g_{ik}\partial_j\xi^k\,,
\end{equation}
and internal $\ast$-automorphisms of fields $X^A$,
\begin{equation}\label{7}
\delta X^A=[\epsilon, X^A]\,,\qquad \delta g_{ij}=0\,,
\end{equation}
where $\xi^i$ and  $\epsilon$ are gauge parameters. In the
commutative limit (i.e. when $\ast$-commutator reduces to the
Poisson bracket) the gauge transformations (\ref{7}) form a subgroup
of the area-preserving diffeomorphisms (\ref{dif}). For this reason
we refer to internal $\ast$-auto\-mor\-phisms~(\ref{7}) as a group of
{\it quantized symplectomorphisms\/}. The latter is also known under
the name of $W_{\infty}$-type symmetry and it has been intensively
studied in the context of conformal field theory and integrable
models~\cite{Pope93,Bakas89}. Thus, we see that although the
commutative limit of our model is equivalent (at least classically) to
the Polyakov string the gauge symmetries underlying both the
formulations are quite different and, in a sense, complementary to each
other: the conformal transformations of the metric rescale the
symplectic structure, while the symplectomorphisms preserve it.

The variation of the action (\ref{af}) leads to the following
equations of motion:
\begin{equation}\label{8}
-\frac{\nu^2}{4}\frac{\delta S}{\delta X^A}=[[X_A,X_B],X^B]=0\,,
\end{equation}
\begin{equation}
\begin{array}{l}\displaystyle
\frac{1}{\sqrt{g}}\frac{\delta S}{\delta g_{kl}}=
-\frac12g^{kl}(\{X^A,X^B\}^2+\alpha^2)+
\nu\left[-\frac18g^{kl}R\{X^A,X^B\}^2+\right.\\
\quad\qquad\qquad\displaystyle
+\frac14\omega^{ik}\omega^{jl}\nabla_i\nabla_j\{X^A,X^B\}^2
-2g^{ki}g^{lj}\{X_A,\{X^A,X^B\}\}\nabla_i\nabla_jX_B+
\\[2mm] \qquad\qquad\quad\displaystyle
+\frac12 g^{kl}\{X_A,\{X^A,X^B\}\}\Box X_B+
+\frac12g^{kl}\{X^A,X^B\}\{X_A,\Box X_B\}-\\[2mm]
\,\,\,\,\qquad\qquad\displaystyle\left.
-\frac12(g^{ik}g^{jl}+g^{il}g^{jk}-g^{ij}g^{kl})
(\nabla_i\{X_A,\{X^A,X^B\}\})\nabla_jX_B\right]+O(\nu^2)=0\,.
\end{array}
\label{9}
\end{equation}
By construction, these equations are known to be invariant under
the $\ast$-automorphisms, but only a part of them is written in a
manifestly covariant form, namely, as double $\ast$-commutator
of~$X$'s. Unfortunately, the compact algebraic representation is
lacking for the variation with respect to metric. It is also seen
that the lowest order of Eqs.~(\ref{9}) determines only the
conformal factor $\sqrt{g}$, not the entire metric $g_{ij}$.

As in the case of undeformed string, the reparametrization invariance of the
model~(\ref{dif}) allows one to bring the metric to the conformally flat
form,
\begin{equation}
g_{ij}=\sqrt{g}\eta_{ij}\,.
\label{10a}
\end{equation}
In this gauge the system of equations (\ref{9}) breaks on two
different parts. The trace part of the variation allows one to
define the conformal factor $\sqrt{g}$ in a pure algebraic way,
expressing it order-in-order via the fields~$X^A$. Substitution of
this expression to the traceless part of the system gives two
constraints on $X^A$ in addition to the equations of motion
(\ref{8}). It is the appearance of these constraints that
drastically differs our model from its matrix counterpart
(\ref{i3}). As it will be clear bellow (see Eqs.~(\ref{12})-(\ref{14a})) they
have the same origin and interpretation as the Virasoro constraints for the
conventional bosonic string coupled to background fields.

Remarkably, the system (\ref{8}), (\ref{9}) admits a special class of
solutions being a deformation of the ordinary string dynamics,
\begin{equation}
\Box X^A=O(\nu^2)\,,\qquad g_{ij}=
\alpha^{-1}\partial_iX^A\partial_jX_A\left(1+\frac{\nu}{4}R\right)+O(\nu^2)\,,
\label{11}
\end{equation}
where $R$ and $\Box$ are the curvature and Laplace operator
associated to the {\it induced\/} world-sheet metric. The
existence of such solutions is provided by the special form of
the first ``quantum correction" to  the undeformed action
functional: it is given by square of undeformed string equations
of motion plus the curvature-depending term, which can be removed
by rescaling of metric.

Let us comment upon the general structure of the deformed string
action.  In any order in $\nu$ it is given by fourth-order
polynomials in the derivatives of the fields $X^A$ contracted
with the help of flat target-space metric. Identifying the
deformation parameter $\nu$ with the reciprocal of string tension
$\alpha^\prime=\alpha/2\pi$ one may treat the deformation as the
interaction vertices for the usual bosonic string moving in a
very special Poincar\'e-invariant background -- all the
background fields are proportional to $\eta_{AB}\eta_{CD}$. It is
not hard to guess about the most general action functional
respecting the gauge transformations~(\ref{dif}),~(\ref{7}) and
leading to a richer content of background fields. It reads
\begin{equation}
\begin{array}{c}\displaystyle
S=S_g+\int{\cal L}_\ast(X)\,,\\ \displaystyle {\cal L}_\ast(X)={\rm d
}\mu \sum\limits_{m=0}^\infty L_{A_1A_2\dots A_m}X^{A_1}\ast
X^{A_2}\ast\dots\ast X^{A_m}\,,
\end{array}
\label{12}
\end{equation}
\enlargethispage{2\baselineskip}
where $L_{A_1A_2\dots A_m}$ are constant tensors and $S_g$ is an
arbitrary scalar functional of metric\footnote{It is pertinent to
note that, living aside $S_g$, the functional (\ref{12}) is
nothing but the action of the general multi-matrix model
\cite{Morozov} written in terms of $\ast$-product and the
respective trace functional. This observation suggests another
link to the $W$-symmetry and related integrable models
\cite{Morozov,int1,int2,Castro00} although we will not dwell on this
point here.}.  Expanding the Lagrangian ${\cal L}_\ast$ in terms of
the derivatives of fields $X^A$, we get
\begin{equation}
\begin{array}{c}\displaystyle
{\cal L}_\ast(X)=\omega \sum G^{(f)}_{A_1\dots
A_m}(X)\nabla_{i(n_1)}X^{A_1}\dots
\nabla_{i(n_m)}X^{A_m}f^{i(n_1)\dots i(n_m)}(R,\nabla R,\dots)\,,\\[2mm]
\displaystyle
\nabla_{i(n_k)}\equiv\nabla_{i_1}\nabla_{i_2}\dots\nabla_{i_{
\scriptstyle n_k}}\,.
\end{array}
\label{13}
\end{equation}
Here tensors $G^{(f)}(X)$ play the role of the background fields
compatible with the fundamental symmetries of the model and their
structure is completely determined by the tensors~$L$.  Passing to
the dimensionless coordinates, $X^A\rightarrow \sqrt
{\alpha'}X^A$, we turn the expansion (\ref{13}) to that in powers of
$\alpha '$ or, what is the same, in a number of derivatives. The
first several terms of the resulting expansion are
\begin{equation}\label{14a}
{\cal L}_{\ast}=\omega
\left(\frac1{\alpha'}V(X)+g^{ij}\partial_iX^A\partial_jX^BG_{AB}(X)
+\omega^{ij}\partial_iX^A\partial_jX^BH_{AB}(X)+D(X)R+O(\alpha')\right).
\end{equation}
The background fields $V,G,H,D$ associated naturally to the
tachyon, metric, antisymmetric tensor and dilaton are not
completely arbitrary -- as they arise from the same set of tensors
$L$ -- but subject to certain constraints. In particular,
\begin{equation}\label{14b}
D(X)=\frac14V(X)=\sum_{k=0}^{\infty} L_{A_1A_2\dots A_k}X^{A_1}X^{A_2}
\dots X^{A_k}\,.
\end{equation}
It would be interesting to find the full system of the
constraints ensuring the invariance of the above action  with
respect to $\ast$-automorphisms  and compare them with the known
low-energy effective equations of motion for the background fields.

The above consideration allows one to treat the deformed string
action as a special highly symmetrical phase of the conventional string
theory rather than a completely new model.
\vspace{-5mm}
\section{Noncommutative string instantons and\protect\\
holomorphic curves}\hspace*{\parindent}
The form of the string action (\ref{af}) is similar to that of
(noncommutative) Yang-Mills theory and  this analogy extends, in a sense, to
their equations of motion: in both cases the immediate search for nontrivial
exact solutions seems to be a hopeless task. The alternative appears,
however, when one studies the Euclidean version of Yang-Mills theory in four
dimensions. Here it is possible  to present the Yang-Mills Lagrangian as the
square of chiral part of the strength tensor and replace the initial
equations of motion by stronger ones -- (anti-)self-duality
conditions.  This is how the remarkable instanton solutions
emerge. In this section we show that the same ``instanton
philosophy'' works well in the case of commutative and
noncommutative string theory providing us with class of
easy-to-construct and study solutions.

We begin with the undeformed string action
\begin{equation}
S=\int\omega\left(\{X^A,X^B\}^2+\frac{\alpha^2}{2}\right)\,,
\label{14}
\end{equation}
with ${\mathbb R}^{\,4}$ being a target space. Notice that the sign
before the last term differs from that in Eq.~(\ref{i1}) due to
the Euclidian signature, otherwise the model has no classical
solutions. The respective equations of motion read
\begin{equation}
\begin{array}{l}\displaystyle
\omega^{ij}\frac{\delta S}{\delta\omega^{ij}}=\{X^A,X^B\}\{X_A,X_B\}-
\frac{\alpha^2}{2}=0\,\\[2mm] \displaystyle
-\frac14\frac{\delta S}{\delta X^A}=\{X^B,\{X_B,X_A\}\}=0\,.
\end{array}
\label{15}
\end{equation}
Given the boundary conditions, the solutions to this system
describe the minimal two-surfaces in ${\mathbb R}^{\,4}$ and a part of
them can be obtained by the same trick one uses to derive the
Yang-Mills instantons. Namely,  we can
rewrite the action~(\ref{14}) in the form
\begin{equation}
\begin{array}{c}\displaystyle
S=\frac12\int\limits_M\omega\left(\{X^A,X^B\}+\alpha^{AB}\right)_\pm^2+
\int\limits_{\partial M}\theta\,,\\ \displaystyle
\theta=2\alpha_\pm^{AB}X_A\partial_iX_Bdl^i\,.
\end{array}
\label{16}
\end{equation}
Here subscripts $\pm$ stand for the (anti-)self-dual  part of a
second-rank skew-symmetric tensor on~${\mathbb R}^{\,4}$,
\begin{equation}
H^{AB}_\pm=H^{AB}\pm\frac12\epsilon^{ABCD}H_{CD}\,,
\label{17}
\end{equation}
$\alpha^{AB}=-\alpha^{BA}$ is a constant tensor normalized as
$(\alpha^{AB}_\pm)^2=\alpha^2$, and  $dl^i$ is a line element on
the boundary $\partial M$.  Because of Euclidian signature we
have the obvious inequality
\begin{equation}
S\geqslant\int\limits_{\partial M}\theta\,,
\label{18}
\end{equation}
and equality is attained iff the following conditions are satisfied:
\begin{equation}
(\{X^A,X^B\}+\alpha^{AB})_\pm=0\,.
\label{19}
\end{equation}
The solutions to these equations provide the minima for the
action functional and may be thought of as string counterparts
for the Yang-Mills instantons. The initial equations of motion~(\ref{15}) are
obvious consequences of the obtained ones.

It is instructive to rewrite the system (\ref{19}) in $\rm
SO(3)$-covariant form
\begin{equation}
\{X^4,X^a\}=\frac12\varepsilon_{abc}\{X^b,X^c\}-\alpha^{4a}_\pm\,,
\qquad a,b,c=1,2,3\,.
\label{20}
\end{equation}
Using the rotation symmetry we bring the vector $\alpha^{4a}$ to the form
$(0,0,\alpha)$. Now let us introduce the complex coordinates in
$\mathbb R^{\,4}=\mathbb C^{\,2}$ by the rule
\begin{equation}
Z=X_1+iX_2\,,\qquad W=X_3+iX_4\,.
\label{21}
\end{equation}
In these coordinates the self-duality equations (\ref{19}) read
\begin{equation}
\{Z,W\}=0\,,\qquad
\{Z,\overline{Z}\}+\{W,\overline{W}\}=2i\alpha\,.
\label{22}
\end{equation}
The first Poisson bracket means that fields $Z(u,v)$ and $W(u,v)$
are functionally dependent, say $W=f(Z)$. Then the second
equation is just a definition of the induced volume form on
the holomorphic curve $Z=f(W)$\,,
\begin{equation}
\omega=(2i\alpha)^{-1}(dZ\wedge d\overline{Z}+dW\wedge
d\overline{W})=
\left(\frac{1+|f'(Z)|^2}{2i\alpha}\right)dZ\wedge d\overline{Z}\,.
 \label{22a}
\end{equation}
In such a way we reproduce the well-known fact from the geometry:
any holomorphic surface in a K\"ahler manifold is globally
minimal. So, the complex geometry delivers a simple method for
generating exact solutions. For example, equation $Z^2=W$ being resolved
in a parametric form, gives
\begin{equation}
X_1=u\,,\qquad X_2=v\,,\qquad X_3=u^2-v^2\,,\qquad X_4=2uv\,,
\label{23}
\end{equation}
that defines  (after compactification of
$\mathbb{R}^{\,4}=\mathbb{C}^{\,2}$ to $\mathbb{C}P^{\,2}$)  the well-known
minimal embedding of two-sphere $S^2\sim \mathbb{C}P^{\,1}$ into
$\mathbb{C}P^{\,2}$ as a projective line. The high-genus minimal
surfaces arise as (hyper)elliptic curves $Z^2=P(W)$, $P(W)$ being a
polynomial without multiple roots.

The generalization of the above constructions to the
noncommutative case is straightforward. Up to an inessential
boundary term the action (\ref{af}) can be  rewritten as
\begin{equation}
S=\frac12\int\limits_M d\mu\,\left([X^A,X^B]+\alpha^{AB}\right)_\pm
\ast\left([X_A,X_B]+\alpha_{AB}\right)_\pm\,.
\label{24}
\end{equation}
Then the global minimum of the functional is achieved provided
\begin{equation}
\left([X^A,X^B]+\alpha^{AB}\right)_\pm=0\,.
\label{25}
\end{equation}
Of course, the transition from (\ref{24}) to (\ref{25}) requires
a justification as we deal now with the $\ast$-square integrated
with respect to the trace measure, and the both are not positive
definite in general. The implication is really correct due to the
following line of reasons. First, we confine ourselves by the
fields to be formal power series in the deformation
parameter~$\nu$ with coefficients being  smooth functions on a
compact manifold~$M$,
\begin{equation}\label{251}
X^A=X^A_0+\nu X^A_1+\nu ^2 X^A_2+\dots\,,
\end{equation}
\begin{equation}\label{25a2}
g^{ij}=g^{ij}_0+\nu g^{ij}_1+\nu ^2g^{ij}_2+\dots\,.
\end{equation}
Then, ignoring a subset of measure zero in $M$, we may assume the
fields to be defined on a closed disc with the Darboux
coordinates $u, v$, so that $\omega = du\wedge dv$. Since,
locally, all the deformation quantizations are known to be
equivalent to each other there exist a formally invertible operator
$B=1+\nu B_1+\nu ^2 B_2+\dots\,$, $B_k$ being finite-order differential
operators, intertwining a given $\ast$-product with the
Weyl-Moyal one \cite{Fedosovbook}. Explicitly,
\begin{equation}\label{25b}
f\ast g= B^{-1}(B(f)\ast _{WM}B(g))\,.
\end{equation}
As this takes place, the trace of $\ast$-product of two functions
is given by the well-known Weyl's formula
\begin{equation}\label{25c}
Tr(f\ast g)=\int dudv B(f)\cdot B(g)\,,
\end{equation}
where $dudv=B^{\ast}d\mu$ is the canonical measure on the disc,
and dot means ordinary multiplication. Now the implication
(\ref{24})$\Rightarrow$(\ref{25}) immediately follows from
nondegeneracy of the operator~$B$.

Rewriting Eqs.\ (\ref{25}) in terms of complex
coordinates~(\ref{21}) we get
\begin{equation}
[Z,W]=0\,,\qquad [Z,\overline{Z}]-[W,\overline{W}]=2\alpha\nu\,.
\label{26}
\end{equation}
Proceeding in a close analogy with above analysis we can solve the
first equation by assuming $Z$ to be a $\ast$-holomorphic
function in $W$,
\begin{equation}
Z=f_\ast(W)=\sum f_n(W\ast)^n\,.
\label{27}
\end{equation}
As to the second (real) equation, it serves then for the
perturbative definition of the metric $g_{ij}$ written in the
conformal gauge (\ref{10a}).

It is pertinent to note that equations of the form (\ref{26})
have also arose as those describing the static BPS membrane
configurations in the BFSS matrix
approximation~\cite{CorTay98,Cor98,CorSchiap99}.
The operator equations of the
form (\ref{25}) have been discussed recently in~\cite{Schwarz01} in
the context of noncommutative Yang-Mills instantons.

In principle, we may forget for the moment about the origin of Eqs.\
(\ref{26}) and treat them as a set of commutation relations for abstract
operators $Z$ and $W$.  From this viewpoint it is clear, that expressions
(\ref{27}) do not exhaust all possible solutions for the commutativity
condition $[Z,W]=0$: there are many commuting operators (and even
matrices), which are not the functions of each other. It is very
intuitive  that solutions of such a type have no definite
``classical limit" as $\nu\to 0$, and hence, poorly fit in a
framework of the deformation quantization. This does not mean that
such non-perturbative solutions  have nothing to do with the
noncommutative string paradigm at all. One of the possible
scenario to reproduce them here is to extend the space of formal
power series~(\ref{251}) to the class of the, so called,
oscillating symbols having the form
\begin{equation}
f(u,v;\nu)\exp\left(\frac{i}{\nu}\varphi (u,v)\right)\,, \label{28}
\end{equation}
where $f=f_0+\nu f_1+\dots\,$ is
a formal power series in the deformation parameter and
$\varphi,f_0,f_1,\dots$ are smooth functions on $M$. Expression (\ref{28})
resembles the WKB-ansatz for the Wigner functions on the phase space and
can be used to approximate the non-perturbative solutions for
Eq.~(\ref{26}). It is these solutions, if any, that one may identify  with
{\it the genuine noncommutative string instantons}.
\subsection*{Acknowledgments}
This work is supported in part by the INTAS under the grant N 00-262
and Russian Ministry of Education under the grant E-00-33-184. The work
of I.V.G. is partially supported by RFBR under the grant N 00-02-17-956.
The work of A.A.S. is partially supported by the RFBR Grant
for Young Scientists N 01-02-06420.

\end{document}